# A Software-defined SoC Memory Bus Bridge Architecture for Disaggregated Computing

Dimitris Syrivelis
IBM Research, Ireland

Andrea Reale
IBM Research, Ireland

Kostas Katrinis
IBM Research, Ireland

Christian Pinto
IBM Research, Ireland

## ABSTRACT

Disaggregation and rack-scale systems have the potential of drastically decreasing TCO and increasing utilization of cloud datacenters, while maintaining performance. While the concept of organising resources in separate pools and interconnecting them together on demand is straightforward, its materialisation can be radically different in terms of performance and scale potential.

In this paper, we present a memory bus bridge architecture which enables communication between 100s of masters and slaves in todays complex multiprocessor SoCs, that are physically intregrated in different chips and even different mainboards. The bridge tightly couples serial transceivers and a circuit network for chip-to-chip transfers. A key property of the proposed bridge architecture is that it is software-defined and thus can be configured at runtime, via a software control plane, to prepare and steer memory access transactions to remote slaves. This is particularly important because it enables datacenter orchestration tools to manage the disaggregated resource allocation. Moreover, we evaluate a bridge prototype we have build for ARM AXI4 memory bus interconnect and we discuss application-level observed performance.

## KEYWORDS

System Bus Interconnect, Software-defined Infrastructures, Dissagregated Computing

## 1 INTRODUCTION

Resource utilization is one of the key performance indicators for internet-scale datacenter and cloud providers to optimize cost of ownership. Guaranteeing consistent high utilization of resources in large datacenters is a daunting task: typical Cloud application mixes show high diversity in terms of their computing resource requirements (i.e., CPUs, memory, storage and accelerators); for example, as studies in [15] and [5] show, the distribution of per-application Memory/CPU usage ratio can be spread over three orders of magnitude.

Modern Cloud implementations largely rely on virtualization and related migration techniques to improve overall datacenter utilization by partitioning and isolating resources of bare metal servers into finer-grained units. However, as virtual machines (VMs) or containers cannot span across the boundaries of a standalone server machine, the overall resource ratio remains constrained to the proportionality imposed by the server mainboard, fixed at datacenter design time. This results in a waste of CPU cores, memory and accelerators when they are asymmetrically depleted.

Fine-grained disaggregation of datacenter resources and their organization into flexible pools has the potential to radically change this landscape. This becomes particularly challenging for tightly integrated components like CPU, memory and coprocessor accelerators that are tied together via a memory bus interconnect or immediately attached to high performing bridges like PCI-e. These components have very high bandwidth and very low latency requirements and also exhibit high variance in the data exchange sizes which cannot be efficiently accommodated by data transport architectures that have been optimized for block transfers (e.g. RDMA over infiniband or converged Ethernet).

In this paper, we explore a different approach and propose a software-defined, SoC memory bus bridge architecture that enables the disaggregation of components that are directly interfaced to the SoC memory bus. The proposed architecture allows local bus masters to communicate with remote slaves that are located on different physical SoCs over a circuit network. In the sequel, we present the memory bridge architectural blueprint and prototype and we discuss the evaluation results of a disaggregated memory case study.

## 2 ARCHITECTURAL BLUEPRINT

The goals of the proposed bridge architecture are:

*(a)* enable computing resources disaggregation at the memory bus level, *(b)* enable deep-software defined support so that resource disaggregation can be controlled by a datacenter resource manager, *(c)* achieve an acceptable tradeoff between performance and scalability and, *(d)* be a generic design that can be used with different memory bus architectures.

Current SoC memory bus architectures [2] [18] provide totally decoupled communication between many masters and slaves. Typically, these architectures feature a number of different parallel channels to concurrently issue read and write transactions. On each and every clock cycle the SoC memory bus masters push data flits on all channels that may belong to different transactions and even be addressed to different slaves. The masters and slaves are memory mapped on the bus in a fixed manner so the physical address of each request is used to identify the recipient master or slave port.

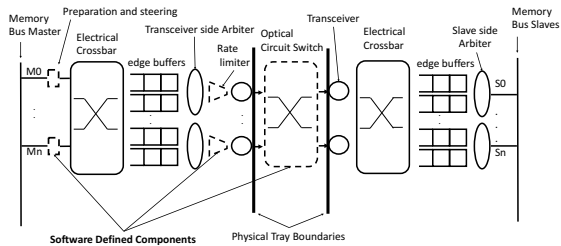

Figure 1: The masters-to-slaves datapath components of the memory bus bridge are depicted. The software defined components are highlighted with dotted lines and bold lines indicate the physical boundaries of the physical component trays.

In figure 1 a high-level design of the proposed bridge architecture is depicted for the many masters to many slaves communication. At the beginning the master bus channels have to be multiplexed in time for streaming communication. In the sequel, the multiplexed channel data flits enter the bridge pipeline which prepares physical addresses and is steering the requests towards the disaggregated memory bus slave. The flits are appropriately demultiplexed at destination.

Since each memory bus instance has it's own fixed address space configuration the bridge end points are appropriately mapped to serve a dedicated address region on each bus side. Evidently, when a memory request is delivered to the local bridge side it is issued to a physical address that is relevant to where the current end-point is mapped. To maximize flexibility, it is not required for the remote slave to be accordingly mapped on the destination bus in order to receive the request with the physical address intact. Instead, the bridge features a request preparation and steering unit that can be configured, at runtime, to further breakdown the address space into small regions which can be handled differently. The handling involves recalculation of the physical address (by applying an appropriate offset) so it can be correctly received by the remote slave on the remote bus map. Subsequently the request flits are sent to the appropriate transceiver that is directly interfaced to the target SoC tray. The memport construct (depicted in 2 is keeping the required data and can be configured by software at runtime.

The serial transceivers lie at the edge of each SoC. Those are expected to get interfaced via one or more circuit switching layers with remote SoCs. The circuit switches should feature a programmable control plane so that they can switch ports at runtime. They can be either electrical crossbars or all optical solutions [13].

To deal with possible asymmetric performance issues, the proposed bridge employs the edge buffering technique and implements software controlled rate limiter support at the master port side. The bridge pipeline assumes that backpressure support is only available up to the serDES pipelines. Once the data flits leave the serDES pipelines towards the circuit network no backpressure support is expected. In addition the bridge architecture considers the circuit network links to be lossless, and does not support any acknowledgement and retransmission scheme.

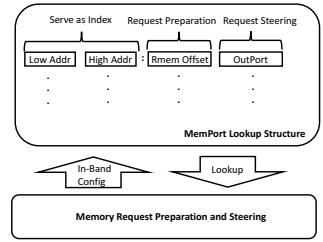

Figure 2: The memport structure (one instance per bus master) hosts information to prepare physical addresses for destination slave and steering towards the designated transceivers. It can be configured from software via an in-band channel at runtime.

Taking advantage of the key features described above, the bridge architecture allows for an entirely cut through and transparent switching design, which just needs to piggy back a few bits of forwarding information on data flits. Moreover, serDES pipelines are completely decoupled to operate in the transceiver clock domain and pull data flits from the edge buffers. Last but not least, appropriate arbiter logic is designed to demultiplex dataflits that arrive from different masters and slaves as well as fairly distribute the available transceiver bandwidth.

## 3 CASE STUDY

A fully functional prototype of the proposed bridge architecture has been realized on reconfigurable logic and has been used for the evaluation of a disaggregated memory system. Disaggregated memory has the potential to change the way the Cloud datacenters are being built because it breaks the fixed ratio between compute and memory resources that is currently imposed by the fixed server trays. The ability to dynamically assign memory resources beyond the traditional server boundaries, allows for more efficient cloud workload allocation that maximizes the datacenter resource utilization [15]. Moreover, todays memory controllers exhibit very low access latencies and bandwidth that linearly scales with the number of cores, so it is a good candidate to assess the proposed bridge performance.

In particular, the Xilinx Ultrascale+ MPSoC platform [17] has been used. The unique feature of this platform is the integration, in a single SoC, of a so-called Processing System (PS) consisting of four ARMv8 A53 cores, and a Programmable Logic (PL) featuring a programmable FPGA. The PS is interacting with the PS side via an implementation of ARM AXI4 interconnect memory bus architecture [18] via 2 master ports, which serve a total of 448GB range and also 2 slaves that are directly interfaced to the PS DDR controller port.

Two MPSoC platforms have been used as depicted in 4 and we interconnected them using 2 GTH transceivers clocked to operate at 10Gb/s over SFP+ copper cables. For the serDES support we have used the Xilinx Aurora protocol in streaming mode. One platform is assigned the role of a compute node and has all the PL master range assigned to the bridge whereas the second is assigned the role of the memory node and interfaces the PS DDR slaves to the bridge. The memory bridge datapath requires 134 cycles for a data flit round-trip which equals to 800$nsecs$ in the current prototype.

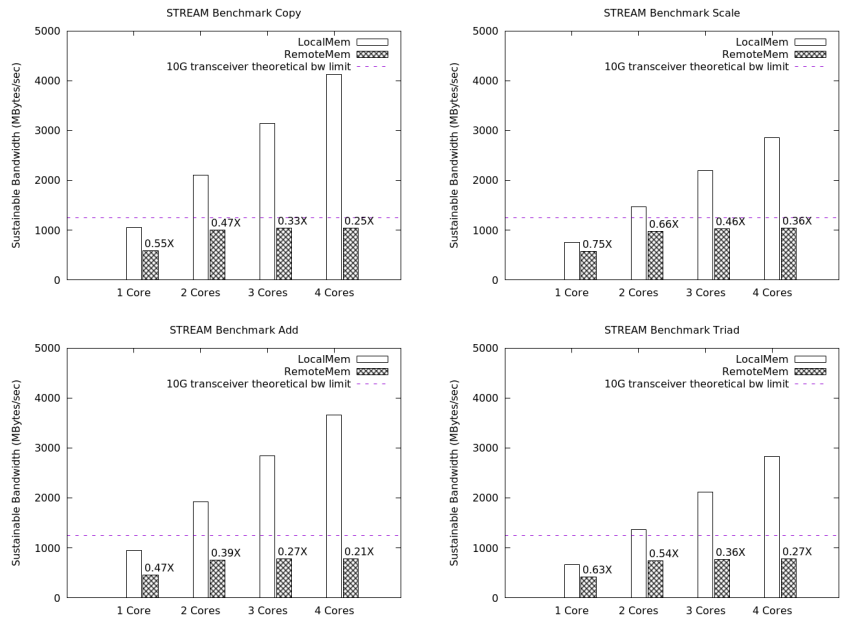

Figure 3: STREAM benchmark performance comparison: local Vs software-defined remote memory

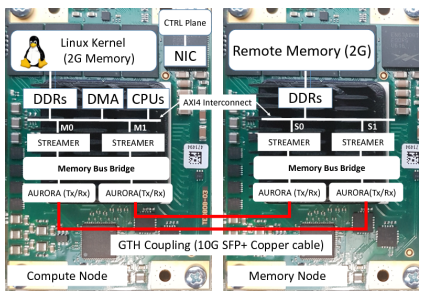

Figure 4: The disaggregated memory prototype

On the software side, we have ported the memory hotplug support [8] for arm64 linux kernel, which enables the dynamic mapping of physical memory segments as paged memory. In addition, taking advantage of the linux kernel Non-uniform memory access support, user applications can explicitly allocate pages from the local or the disaggregated memory pool. Simple orchestration control has been implemented to configure the bridge datapath to accordingly map memory segments and compute memory offsets for the experiments that follow.

We evaluate the disaggregated memory performance with the STREAM benchmark [9], the de facto industry standard to measure sustainable memory bandwidth and overall processing balance as perceived by user space applications. We configured STREAM to use 10 million array elements, requiring a total memory of 228.9 MiB, which is well beyond the MPSoC cache size. Each benchmark run executes four kernels, i.e., "copy", "scale", "sum" and "triad" [9]. Specifically, "copy" reads/writes 16 bytes (1 read, 1 write ops) of memory per iteration, performing no floating point operations (FLOPs); "scale" reads/writes the same amount of memory with the same number of operations but it performs 1 FLOP per iteration; "sum" accesses 24 bytes of memory (2 read and 1 write ops) and executes 1 FLOP per iteration; finally, "triad" accesses 24 bytes of memory (2 read and 1 write ops) executing 2 FLOPs per iteration. Using the OpenMP support built-in on STREAM, we confine the benchmarks to run on 1 up to all 4 compute node cores. Leveraging the local and remote NUMA domains, we repeat the same executions using only local or disaggregated memory.

Figure 3 shows the results of our evaluation, comparing local and remote memory performance through clustered bars. The dotted line designates the maximum theoretical bandwidth that can be achieved by a 10G transceiver, i.e., 1280 MiB/s and the bars the sustainable memory bandwidth as measured by STREAM benchmark for a different number of cores.

Focusing on the "copy" kernel, the results show that one CPU core can achieve 562 MiB/s bandwidth towards remote memory, with a penalty of 47% compared to local access. As more CPUs are used concurrently, the transceiver bandwidth is quickly saturated and, beyond 2 CPUs, it becomes the performance bottleneck. In terms of absolute bandwidth, the "scale" benchmark has worst results because of the presence of the additional FLOP. However, when comparing local vs. disaggregated memory, the application-perceived penalty of using remote rather than local memory is reduced to 25%, due to the more balanced mix of memory access / processing operations. The same trend can be observed in the "add" (24 bytes memory accesses per iteration, with 1 FLOP) and "triad" (24 bytes memory accesses per iteration, with 2 FLOPs). Overall, these results validate the balanced, pipelined design and implementation of the proposed memory bus bridge, and demonstrate that is capable of exploiting the full potential of the AXI4 interconnect parallel and asynchronous operation.

## 4 RELATED ART

Exporting the memory bus to MPSoC off-chip components is critical for flexible and high performance computing, and commercial grade solutions like OpenCAPI [3], HyperTransport [1] and QuickPath Interconnect [7] are nowadays available. These approaches primarily target components on the same tray and require fixed configuration that cannot be changed at runtime. Therefore they have not been considered for datacenter scale disaggregation support.

Making native datacenter resource pooling a reality in the datacenter through disaggregation has been a year-long quest. Breaking the monolithic design of datacenters (including memory) to decouple arbitrary workload sizes from static server configuration and enabling independent technology refreshes of various components has been one of the missions of the Open Compute Project [14] since its early days. Notable demonstrators and prototype concepts include Intel Rack Scale Design [4], Facebook "Group Hug" and "Yosemite" server designs, as well as production-grade specialized kernels and platform orchestration software for virtual machines operating on pooled servers, such as Liqid [6]. Similarly, the HPE Machine [12] prototype showcases SoCs accessing remote memory via specialized bridging controllers and fabric. Our work shares common objectives with and can act complementary to such and related designs; our unique ambition and the main differentiation point of our proposal stands in its ability to offer disaggregated access dynamically and transparently to unmodified masters and slaves and at memory-scale performance levels.

Our architecture shares concepts with the approach presented in [10] for disaggregated memory. To maintain a simple hardware controller design, requiring only a few configuration registers, the authors prefer to integrate all the remote memory forwarding information in the physical address and exploit the OS page translation infrastructure for accessing remote memory. This dependency on the OS page mechanism does not allow the proposed architecture to be used in a wider context to integrate accelerators or other types of slaves that are not acting as system memory.

Shared memory clusters define a machine organization taxonomy that shares technical challenges and some of the technical and business objectives of disaggregated datacenters. Distributed shared-memory machines representative of this taxonomy, like NumaScale [11] or SGI UV [16], have emerged with the goal of satisfying parallel and distributed applications (e.g., large-scale computational science, mainframe computations) that require deployment on a large number of tightly cooperating cores, whereby also in-memory computing can bring substantial benefit.

## 5 CONCLUSIONS AND FUTURE WORK

In this paper, we have presented a software-defined memory bus bridge architecture for fine-grained components disaggregation in warehouse-scale computer and cloud datacenters. Bridging memory bus requests between different SoCs and export deep software-defined support to orchestration tools, enables the organization of resources in independent pools that can be dynamically wired at runtime to form computing platforms.

With the advancement in serDES latency and optical circuit switch performance, high speed interconnects like the SoC memory bus can now issue transactions over the network to remote slave peripherals, in a transparent manner, as if they were locally attached. While the remote access performance is no match compared to the local access delays and bandwidth, we have demonstrated via our prototype evaluation that acceptable levels can be achieved, even for the straightforward disaggregation of main system memory.

We are currently advancing our architecture to handle interrupt delivery so we can support the disaggregation of all peripheral types and memory-mapped bridges like the PCI-e, as well as to provide appropriate mailbox support that will enable disaggregated paravirtualization.

## ACKNOWLEDGEMENTS

This work has been conducted in the scope of the dReDBox (disaggregated Recursive Datacenter-in-a-Box) project, which has been funded by EU Horizon 2020 Research and Innovation programme under grant agreement No 687632.

## REFERENCES
[1] AMD. Hypertransport interconnect. Online: https://hypertransport.org.
[2] R. A. Bergamaschi and W. R. Lee. Designing systems-on-chip using cores. In *Proceedings of the 37th Annual Design Automation Conference*, DAC '00, pages 420–425, New York, NY, USA, 2000. ACM.
[3] O. Consortium. Opencapi. Online: http://opencapi.org.
[4] I. Corp. Intel rack scale design. Online: http://www.intel.com/content/www/us/en/architecture-and-technology/rack-scale-design/rsd-vision-brochure.html.
[5] S. Han, N. Egi, A. Panda, S. Ratnasamy, G. Shi, and S. Shenker. Network support for resource disaggregation in next-generation datacenters. In *Proceedings of the Twelfth ACM Workshop on Hot Topics in Networks*, HotNets-XII, pages 10:1–10:7, New York, NY, USA, 2013. ACM.
[6] L. Inc. Liqid hyperkernel. Online: https://liqid.com.
[7] Intel. Quickpath interconnect. Online: https://www.intel.com/content/www/us/en/io/quickpath-technology/quickpath-technology-general.html.
[8] kernel.org. Linux memory hotplug documentation. Online: https://www.kernel.org/doc/Documentation/memory-hotplug.txt.
[9] J. D. McCalpin. Memory bandwidth and machine balance in current high performance computers. *IEEE Computer Society Technical Committee on Computer Architecture (TCCA) Newsletter*, pages 19–25, Dec. 1995.
[10] H. Montaner, F. Silla, H. Fröning, and J. Duato. A new degree of freedom for memory allocation in clusters. *Cluster Computing*, 15(2):101–123, 2012.
[11] Numascale AS. Numaconnect: A high level technical overview of the numaconnect technology and products (numascale whitepaper). Online: https://www.numascale.com/numa_pdfs/numaconnect-white-paper.pdf/.
[12] T. N. Platform. Hpe powers up the machine architecture. Online: https://www.nextplatform.com/2017/01/09/hpe-powers-machine-architecture/.
[13] Polatis. *Polatis all-optical SDN switches series 7000*.
[14] O. C. Project. Ocp summit iv: Breaking up the monolith. Online: http://www.opencompute.org/blog/ocp-summit-iv-breaking-up-the-monolith/.
[15] C. Reiss, A. Tumanov, G. R. Ganger, R. H. Katz, and M. A. Kozuch. Heterogeneity and dynamicity of clouds at scale: Google trace analysis. In *ACM Symposium on Cloud Computing (SoCC)*, San Jose, CA, USA, Oct. 2012.
[16] SGI UV. The world most powerful in-memory supercomputers. Online: http://www.sgi.com/products/servers/uv/index.html.
[17] Trenz Electronic. Trenz UltraSOM+ TE0808-04. Online: https://shop.trenz-electronic.de/en/TE0808-04-09-S-TE0808-04-09-S-Starter-Kit?c=329b.
[18] Xilinx. *AXI Reference Guide v13.1 UG761*, March 2011.